\documentclass[conference]{IEEEtran}
\input epsf
\usepackage{graphicx}
\usepackage{listings}
\usepackage{xcolor}
\usepackage{amsmath}

\definecolor{codegreen}{rgb}{0,0.6,0}
\definecolor{codegray}{rgb}{0.5,0.5,0.5}
\definecolor{codepurple}{rgb}{0.58,0,0.82}
\definecolor{backcolour}{rgb}{0.95,0.95,0.92}

\lstdefinestyle{mystyle}{
    backgroundcolor=\color{backcolour},   
    commentstyle=\color{codegreen},
    keywordstyle=\color{magenta},
    numberstyle=\tiny\color{codegray},
    stringstyle=\color{codepurple},
    basicstyle=\ttfamily\footnotesize,
    breakatwhitespace=false,         
    breaklines=true,                 
    captionpos=b,                    
    keepspaces=true,                 
    numbers=left,                    
    numbersep=5pt,                  
    showspaces=false,                
    showstringspaces=false,
    showtabs=false,                  
    tabsize=2
}
\usepackage[a4paper, margin=1in]{geometry}

\begin{document}
\title{\LARGE Implementing PID Controller on a hand follower robot using Numerical Methods}

\author{\IEEEauthorblockN{ Mostafa A. Mostafa (1), Abdallah A. Mohamed (1), Ahmed H. Aborehab (1),\\ Mohamed A. Shabaan (1), Mohamed K. Mohamed (1),\\ Mostafa O. Mohamed (1), Tarek M. Shohdy (1)}
\IEEEauthorblockA { (1) Egypt-Japan University of Science and Technology, Egypt}}

\maketitle

\begin{abstract}
Overall, in any system, the proportional term, integral term, and derivative term combined to produce a fast response time, less overshoot, no oscillations, increased stability, and no steady-state errors. Eliminating the steady state errors connected to typical PID systems is crucial for achieving stability. To plot the transfer function's responses with various integrator gains for auto tuning, a MATLAB M-file was developed. Auto tuning techniques were then applied to PID systems to eliminate steady state defects.
this paper analyzes and tests the improvement of PID controller over the regular P-controller taking a hand follower robot as a system example using methods with
simulation and numerical analysis study.
\end{abstract}

\IEEEoverridecommandlockouts

\IEEEpeerreviewmaketitle
\section{Introduction}
Proportional-Integral-Derivative (PID) control is the most widely used control algorithm in industry and has become the mainstream of industrial control. The popularity of PID controllers is due in part to their robust performance over a wide range of operating conditions and in part to their functional simplicity, which allows engineers to operate them in a simple and straightforward manner. the settings for this controller should be known to tune it correctly to produce the preferred output. Here, tuning is the process of obtaining the ideal response of the controller by setting the optimum proportional gain, integral and derivative factors.
One type of action used in PID controllers is proportional control. Proportional control is a form of feedback control. It is the simplest form of continuous control that can be used in closed loop control systems. Pure P control minimizes process variable fluctuations but does not always bring the system to the desired set point. It provides a faster response than most other controllers, which makes the P-only controller initially responsive a few seconds faster. However, as the system becomes more complex (i.e. more complex algorithms), the response time differences may accumulate such that the P-controller may respond even minutes faster. Although a pure P controller has the advantage of a faster response time, it introduces deviations from the setpoint. This deviation is called offset and is usually not needed in the process. The presence of an offset means that the system cannot maintain the desired set point in steady state. It is analogous to bias in a calibration curve, where there is always a fixed, constant error that prevents the line from passing through the origin. The offset can be minimized by combining pure P control with another form of control, such as B. I or D control, is combined. However, it is important to note that it is not possible to eliminate the offset.
\cite{chapraapplied}

\begin{figure}[!p!h]
    \centering
    \includegraphics[width = 0.4 \textwidth]{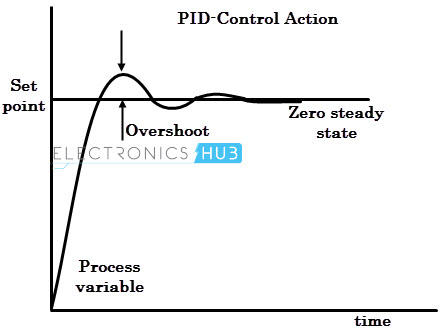}.
    \caption{PID control Action}
    \label{fig:kimenetek.png}
    
\end{figure}

This type of PID controller combines proportional control with integral and derivative control to automatically help equipment compensates for changes within the system. These modifications, integrals, and derivatives are expressed in time-based units. The integral function eliminates offset by summing up the input signal over time keeping running total so it has the memory of what happened before or the past information on other meaning when the system being in steady state under the desired point it creates a nonzero error so the integration will increase the output value, after a few iterations of increasing the output overshooting over the desired value at this point(the point above the desired value) utility of the derivative mode comes to provide fast response and prevent overshooting by calculating the rate of change of how the error growing or shrinking to provide the future information to the system and take a decision for next move depends on that.  However, PID controllers are difficult to tune. but when properly tuned, provide the best control system.
\cite{steven2007applied}

\begin{figure}[!p!t]
    \centering
    \includegraphics[width = 0.4 \textwidth]{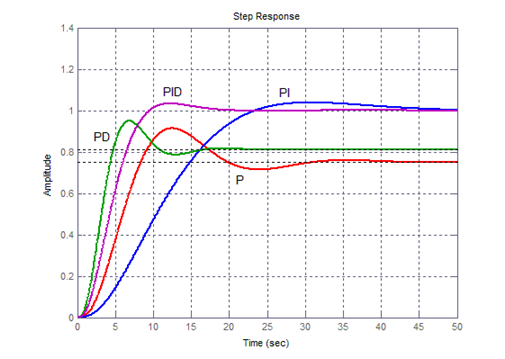}.
    \caption{PID controller output signal}
    \label{fig:kimenetek2.png}
    
\end{figure}
 
\newpage

\section{Objectives}
\begin{itemize}
\item Apply the numerical methods of differentiation in the part of the “D” (differential) in the PID controller which anticipates the future errors.
\item Reflect our studies in the integration using numerical methods in the “I” (Integral) part of the PID controller that calculates the past errors.
\item Keep the robot at the same distance from the target with the highest accuracy possible using the numerical methods studied in different lessons.
\end{itemize}

\section{Analysis}  
In this section, we are going to explain the math behind the PID controller and how we are going to numerically approximate the integral and differential values. Also, the MATLAB implementation for the controller on a differential drive robot to keep him on a certain distance from a wall, object, hand, etc...
\vspace{5mm}
\subsection{Theory}
First, we have to see the mathematical formula for PID controller. For an Error function given by: 
\begin{equation}\label{eq: error eq}
    E(t) =  R - x(t)
\end{equation}
where $E(t)$ is the Error function of the system, $x(t)$ is the distance the robot moved at time $t$, $R$ is the reference point (wall, hand, ...). Reference $R$ can be constant like the case we are studying or can be also a function of $t$ Example: $sin(t)$, $cos(t)$, $e^t$.

The input for the PID controller is the Error function value at time $t$, and then we use the integral of the function to calculate the previous error and compensate that error in the form of overshooting from the desired point, to decrease this overshoot, we have to use the derivative of the Error function to make the output speed starts to decrease as we are approaching the desired point. At the end, can get the linear speed of the robot as:
\begin{equation} \label{eq: PID}
    v = K_p E(t) + K_i \int_{0}^{t} E(t) \,dt + K_d \frac{dE(t)}{dt}
\end{equation}
where $K_p$, $K_i$, $K_d$ are parameters to be tuned to get the best response from the robot.

\vspace*{5mm}

\textbf{1- Evaluating definite Integral using Simpson's rule:}\cite{mckeeman1962algorithm}
Using Simpson's rule to evaluate the definite integral of the function. As $E(t)$ values are recorded each $h$ sec, so we can apply the rule to evaluate the integral $I = \int_{0}^{t} E(t)$:
\begin{equation} \label{eq: Simpson}
    I = \frac{h}{3} [E_1 + 2 \sum_{i=3}^{m-1} E_{2i-1} + 2 \sum_{i=2}^{m} E_{2i} + E_{2m}]
\end{equation}
Where $h$ is the time step size, $i = 1,2,3,...,n$, $n =2m$ and is equal to the number of points we have.
The evaluated integral have a truncation error $O(h^5)$ which is accurate enough for our application.

For even number of points, we cannot apply Simpson's rule, so we have to evaluate Simpson integral for $n-1$ points and the last 2 points to be evaluated using trapezoidal area rule, then we can evaluate $I$ as:
\begin{equation} \label{eq: trapezoidal}
    I = I_{simpson} + \frac{E_n + E_{n-1}}{2} h
\end{equation}

\vspace*{5mm}

\textbf{2- Evaluating Derivative using backward difference:} \cite{matsuno1966numerical}
Primary task for our problem is to get the derivative of the last Error value, we have to use the backward difference method between the last to points:
\begin{equation}
    E'(t) = \frac{E_n-E_{n-1}}{t_n-t_{n-1}}
\end{equation}
but, also we have to consider the accuracy and truncation error of the method used. For the backward difference the truncation error is $O(h)$ which is significant for our problem. However, we can handle this problem by evaluating a form for derivative with higher accuracy using the last 3 points. The Taylor expansion for the Error function of $E(t_{n-1}) = E(t - h)$ and $E(t_{n-2}) = E(t - 2h)$ is:
\begin{equation}
    E(t-h) = E(t) - hE'(t) + \frac{h^2}{2}E''(t) + \frac{h^3}{6}E'''(t)
    \label{eq: taylor_E(t-h)}
\end{equation}
\begin{equation}
    E(t-2h) = E(t) - 2hE'(t) + {2h^2}E''(t) + \frac{8h^3}{6}E'''(t)
    \label{eq: taylor_E(t-2h)}
\end{equation}
By summation of equations \ref{eq: taylor_E(t-h)},\ref{eq: taylor_E(t-2h)}, we can get rid of the $h^2$ term and decrease the truncation error.
\begin{equation} \label{eq: taylor_2_expansion}
\begin{aligned}
    -4E(t-h) & = -4E(t) + 4hE'(t) \\ 
    & \quad - 2h^2E''(t) - \frac{4h^3}{6}E'''(t), \\
    E(t-2h) & = E(t) - 2hE'(t) + 2h^2E''(t) \\ 
    & \quad + \frac{8h^3}{6}E'''(t)
\end{aligned}
\end{equation}
The 2 equations in \ref{eq: taylor_2_expansion} add up to formulate:
\begin{equation}
    E(t-2h) - 4E(t-h) = -3E(t) + 2hE'(t) + \!\frac{4h^3}{6}E'''(t)    
\end{equation}
Finally, we can approximate $E'(t)$ as:
\begin{equation} \label{eq: backward_diff}
\begin{split}
    E'(t) &= \!\frac{E(t-2h) - 4E(t-h) + 3E(t)}{2h} - \!\frac{h^2}{3}E'''(t) \\
    E'(t) &= \frac{E(t-2h) - 4E(t-h) + 3E(t)}{2h} + O(h^2) \\
\end{split}    
\end{equation}
In terms of the error values, we can say that for $n$ points, the backward difference equation is:
\begin{equation} \label{eq: final_backward_diff}
    E'_n = \frac{E_{n-2} - 4E_{n-1} + 3E_{n}}{2h}
\end{equation}

\vspace{5mm}

\subsection{MATLAB Implementation}
 After we understood the mathematical concept for the PID controller and how to interpret the values of the integral and the derivative, now we can implement all the work as MATLAB codes to simulate and test the controller.

\begin{figure}[!p!h]
    \centering
    \includegraphics[width = 0.4 \textwidth]{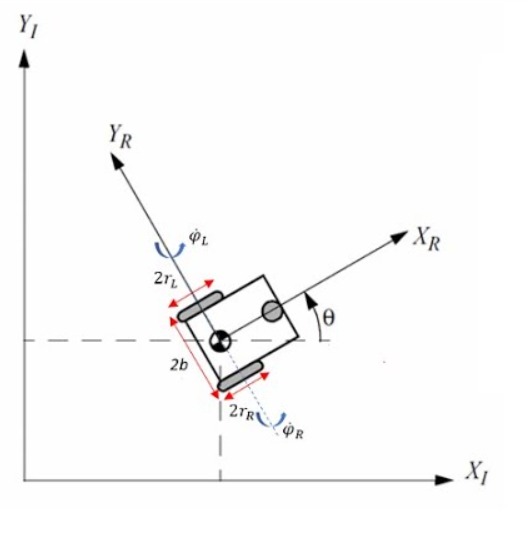}.
    \caption{Differential drive robot}
    \label{fig: diff_drive_robot}
\end{figure}

 First, we must consider the kinematic model for the differential drive robot we are studying. The kinematic model for the robot is
 \begin{equation} \label{eq: diff_kinematics_matrix}
    \begin{bmatrix}
        \Dot{x}\\
        \Dot{y}\\
        \Dot{\theta}
    \end{bmatrix} =
    \begin{bmatrix}
        \frac{r}{2} cos(\theta) & \frac{r}{2} cos(\theta) \\
        \frac{r}{2} sin(\theta) & \frac{r}{2} sin(\theta) \\
        \frac{r}{2b} & \frac{-r}{2b}
    \end{bmatrix}
    \begin{bmatrix}
        \Dot{\phi_r}\\
        \Dot{\phi_l}
    \end{bmatrix}
 \end{equation}
 where $\Dot{x}$, $\Dot{y}$, $\Dot{\theta}$ are the velocities of the robot, $\theta$ is the orientation of the robot from the $x$-axis, $\phi_r$ and $\phi_l$ are the angular velocities of the wheels, $r$ is the radius of the wheels and $b$ is the track width (half the distance between the 2 wheels).
 All as shown in fig. \ref{fig: diff_drive_robot}.
As we just need the $x$ position - the robot will move just forward and backward so the $y$ position and $\theta$ is  - so we can summarize equation \ref{eq: diff_kinematics_matrix} into:
\begin{equation}
    x_{n+1} = x_n + R\Dot{\phi}t
\end{equation}
At the end, we add all the x values of the robot into array and return the array $x$
\begin{equation}
    x = [x_1, x_2, x_3, ..., x_n ]
\end{equation}
Following code \ref{code: MATLAB Robot Kinematic model} is the MATLAB code for the kinematic model 
\begin{lstlisting}[language=Matlab, caption=Differential Drive kinematic model MATLAB code, label = code: MATLAB Robot Kinematic model]
function x = Robot_Kinematics(w,x,t)
% x is an array for all the distances the robot travelled
% w is the angular velocity of the wheels in rad/sec
% t is the time step size
% r = 0.05
    x = [x  x(end) + w*0.05*t];
\end{lstlisting} 

Next, we are going to implement the Simpson's rule in MATLAB. We check first is the number of points odd or even, to classify whether to use Simpson rule equation \ref{eq: Simpson} or Simpson and trapezoidal rule equation \ref{eq: trapezoidal}. Applying the rule, we can use a for loop to sum all the intervals and then multiply by $\frac{h}{3}$. Code represents the MATLAB code for Simpsons's rule.
\begin{lstlisting}[language=Matlab, caption=Simpson Integral MATLAB code, label = code: MATLAB Simpson Integral]
function integral_x = Simpson_integral(x,t)
% x is the array of Error values
% t is the array of time
    if length(t) == 1
        % if the time array is still have only one value return 0
        integral_x = 0;
        return
    end
    h = t(2)-t(1); % h value
    % for odd number of points, even number of intervals we apply simpson
    if mod(length(t),2) == 1
        integral_x = x(1) + x(length(x)); % add first and last value
        for i = 2:length(x)-1
            if mod(i,2) == 1
                % for odd points, we multiply by 2
                integral_x = integral_x + 2*x(i);
            else
                % for even points, we multiply by 4
                integral_x = integral_x + 4*x(i);
            end
        end
        integral_x = integral_x * h/3;
        % multiply by h/3 and return the value
    else
        % for even points, odd number of points we cannot apply only simpson, but take the last 2 points and apply trapezoidal + Simpson
        integral_x = x(1) + x(length(x)-1);
        for i = 2:length(x)-2
            if mod(i,2) == 1
                integral_x = integral_x + 2*x(i);
            else
                integral_x = integral_x + 4*x(i);
            end
        end
        integral_x = integral_x * h/3 + h/2 * (x(length(x)-1) + x(length(x)));
        % trapeziodal rule
    end
\end{lstlisting}

Then, the implementation of the derivative using the 3 points backward difference method. Applying equation \ref{eq: final_backward_diff} on the MATLAB simply by calling the last 3 points from the Error array and dividing by $2h$. Code \ref{code: MATLAB Derivative backward difference} is the implementation of the method for derivative.
\begin{lstlisting}[language=Matlab, caption=Derivative backward difference MATLAB code, label = code: MATLAB Derivative backward difference]
function dx = Derivative_Backward_difference(x,t)
% x is the array of Error values
% t is the array of time
    if length(x) == 1
        % if there is only 1 point return 0
        dx = 0;
        return
    end
    if length(x) == 2
        % for only 2 points apply normal 2 points slope
        dx = (x(2)-x(1))/(t(2)-t(1));
        return
    end
    % return dx vale for by using 3 point backward difference
    dx = (x(end-2) -4*x(end-1) + 3*x(end))/(2*(t(end) - t(end-1)));
\end{lstlisting}

\vspace{5mm}
PID controller depend on the last 2 codes and how accurately they evaluate the integral and derivative, apparently all we need is to apply equation \ref{eq: PID} and give the main function the parameters $K_p$, $K_i$, $K_d$. Code \ref{code: MATLAB PID controller} is the MATLAB function takes input the parameters and error array and time array to pass them to integral and derivative functions.
\begin{lstlisting}[language=Matlab, caption=PID controller MATLAB code, label = code: MATLAB PID controller]
function output = PID_Numerical(E,t,Kp,Ki,Kd)
% Kp, Ki, Kd are tuning parameters for the controller
% E is the array of Error values
% t is the array of time
    output = Kp*E(end) + Ki*Simpson_integral(E,t) + Kd*Derivative_Backward_difference(E,t);
\end{lstlisting}

\vspace{5mm}
Finally, we can put all into work to get the whole system code \ref{code: MATLAB system simulation}. We have to define the time array $t$ and the $h$ for the system, then the initial position and the reference we are trying to reach. Subsequently, we define the tuning parameters $K_p$, $K_i$, $K_d$. 

Last, the for loop that execute the system, as first we build the error array by subtracting the reference from the position values equation \ref{eq: error eq}, pass the error array and the tuning parameters to the PID controller and get the output velocity and pass it to the kinematic model to get the new position array $x$ and loop for the time span we have defined. At the end we plot the $x$ vs $t$ and fine tune the controller using induction.
\begin{lstlisting}[language=Matlab, caption=system simulation MATLAB code, label = code: MATLAB system simulation]
% defining time array and define h
t = 0:0.01:30;
% define the reference and x initial = 0
reference = 1;
x = 0;
% define the tuning parameters
Kp = 10.8;
Ki = 17.7;
Kd = 3.2;
% the loop that excute and simulate the control system 
for i = 2:length(t)
    % E is the error array obtained by subtracting all the x values from the reference
    E = reference - x;
    % w is the velocity of the robot obtained by the PID controller
    w = PID_Numerical(E,t(1:i-1),Kp,Ki,Kd);
    % the new value of x from the kinematic model
    x = Robot_Kinematics(w,x,0.02);
end
% plotting the distance the robot moved fine tune the parameters by induction
plot(t,x)
title("PID controller on Follower Robot using numerical methods")
hold on
plot(t,reference*t./t,'LineStyle','--')
legend('x(t)','Reference')
hold off
\end{lstlisting}

\section{Results}
\begin{figure}[!p!h]
	\centering
	\includegraphics[width = 0.4 \textwidth]{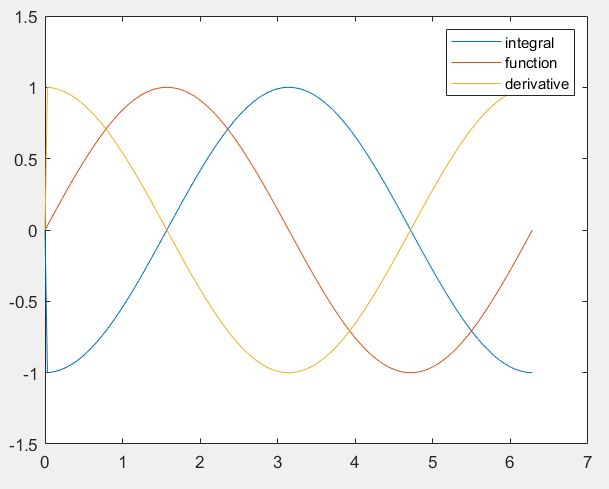}.
	\caption{Derivative and Integral test for sin(x)}
	\label{fig: check}
\end{figure}

\begin{figure}[h]
	\centering
	\includegraphics[width = 0.4 \textwidth]{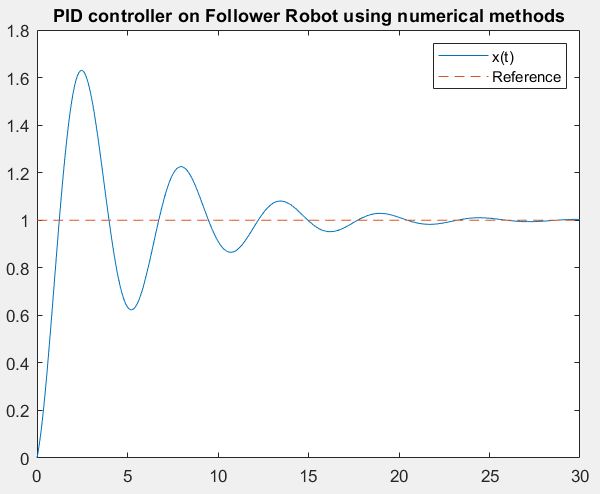}.
	\caption{Controller output signal at KP = 5, KI = 17.7 and KD = 3.2}
	\label{fig: KP}
\end{figure}

\begin{figure}[h]
	\centering
	\includegraphics[width = 0.4 \textwidth]{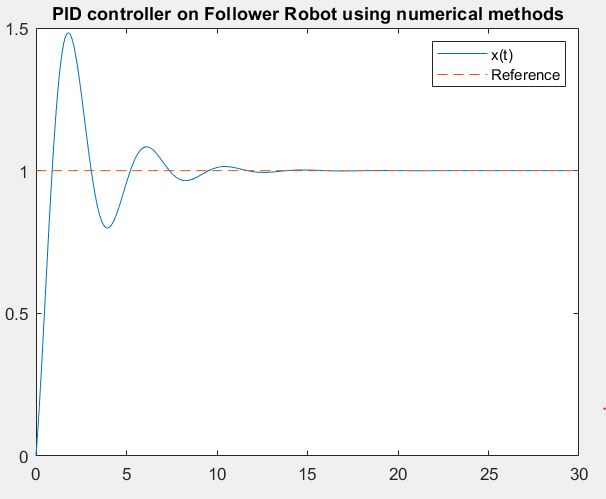}.
	\caption{Controller output signal at KP = 10.8, KI = 30 and KD = 3.2}
	\label{fig: KI}
\end{figure}

\begin{figure}[h]
	\centering
	\includegraphics[width = 0.29 \textwidth]{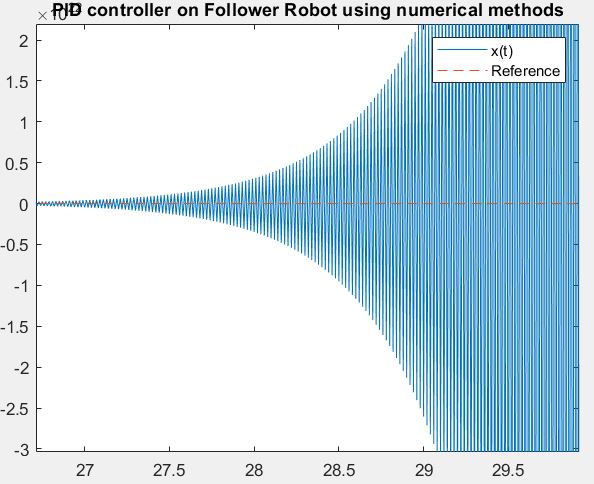}.
	\caption{Unstable output signal at KP = 10.8, KI = 17.7 and KD = 5.1}
	\label{fig: KD}
\end{figure}

\begin{figure}[h]
	\centering
	\includegraphics[width = 0.4 \textwidth]{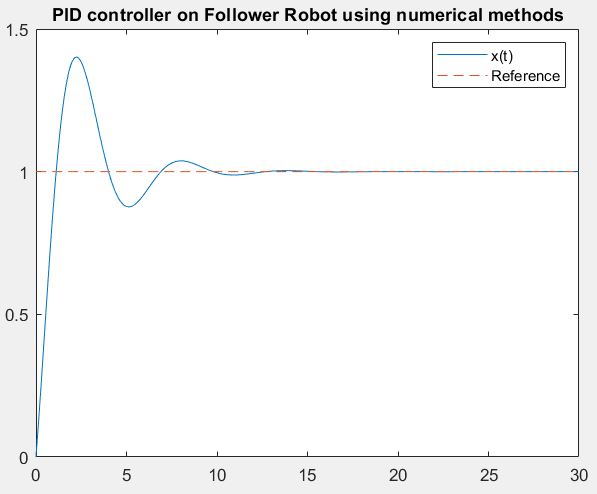}.
	\caption{Controller output signal at KP = 10.8, KI = 17.7 and KD = 3.2}
	\label{fig: Final}
\end{figure}

To validate our work we had to run multiple tests, observe their results and try to give an explanation to what we have observed. This section gives the test results of all the functions and parameters we tested.\\
First of all we tested the derivative and integration functions we implemented on MATLAB to make sure they are giving true answers when we input common functions like trigonometric functions, and the two functions proven success and accuracy as shown in fig. \ref{fig: check} where we tried to get the derivative and integral of sin(x).
 \\
Then we started changing the KP, KI and KD parameters to see how they affect the output signal. Starting with the KP when we kept the KI at 17.7 and KD at 3.2 we started with a KP of 5 which made great oscillation and increased the settling time as shown in fig. \ref{fig: KP}. While increasing the KP we noticed that the controller is more aggressive, the oscillation and the settling time is smaller. We settled on a KP value of 10.8 at last with a conclusion that KP accounts for instantaneous deviation of the set point.
\\After that we left the KP at 10.8 and KD at 3.2 and started to change the KI and observe the output signal. We noticed that increasing the KI eliminates the steady-state error but increases overshoot and oscillations as  in fig. \ref{fig: KI} where we set the KI = 30. At last we found that a KI value of 17.7 is descent and gives a satisfying output signal.
\\Finally we set the KP at 10.8 and KI at 17.7 and started changing the KD value while observing the output signal. when we increased the KD value over 4.9 we noticed that a very high oscillation occurs and the signal does not reach the set point over time. Increasing the KD makes the oscillation more smooth, however high KD value -above 4.9 in our example- can make the system unstable as shown in fig. \ref{fig: KD}. 
\\At last we settled on a KD value of 3.2 to get the controller output signal shown in fig. \ref{fig: Final} by that we settle on the set point in a good time and the oscillation is not considered high.

\section{Conclusion}
As a conclusion for our progress in this report, we applied our knowledge and study of the applied numerical methods to perform a PID controller. So, we used Simpson's rule in order to evaluate our accumulative integral error. Also, we used the backward difference method with two points to evaluate the differential accumulative error. We used this produced PID system to tune a mobile follower robot's motion to keep it at specific distance from an object -can be human, wall or even an obstacle-. As shown above in our analysis and results, we achieved our goal to tune the motion as accurate as possible.

\section{Acknowledgment}
In this section, we can't deny the external efforts that help us a lot in the brainstorming and the implementation phases. We would to thank Prof. Waheed Zahra for his excellent performance through teaching us different numerical methods for solving the differentiation and integration problems and also his effort in the lectures to fetch project ideas for us. We can't deny the Robotics Club efforts for supplying us with the follower robot made by members in the technical committee.


\begin{thebibliography}{99}
\bibitem{chapraapplied}
S.~C. Chapra,  
\newblock \textit{Applied Numerical Methods with MATLAB for Engineers and Scientists},  
\newblock McGraw-Hill International Edition, 2005.

\bibitem{steven2007applied}
S.~C. Chapra,  
\newblock \textit{Applied Numerical Methods with MATLAB: For Engineers and Scientists},  
\newblock Tata McGraw Hill Education Private Limited, 2007.

\bibitem{matsuno1966numerical}
T.~Matsuno,  
\newblock "Numerical integrations of the primitive equations by a simulated backward difference method,"  
\newblock \textit{Journal of the Meteorological Society of Japan. Ser. II},  
\newblock vol. 44, no. 1, pp. 76--84, 1966.

\bibitem{mckeeman1962algorithm}
W.~M. McKeeman,  
\newblock "Algorithm 145: Adaptive numerical integration by Simpson's rule,"  
\newblock \textit{Communications of the ACM},  
\newblock vol. 5, no. 12, p. 604, 1962. 
\end{thebibliography}
\end{document}